\documentclass[conference]{IEEEtran}

\IEEEoverridecommandlockouts
\usepackage{cite}
\usepackage{amsmath,amssymb,amsfonts,bm,bbm}
\usepackage{algorithmic}
\usepackage{algorithm}
\usepackage{setspace}
\usepackage{graphicx}
\usepackage{textcomp}
\usepackage{xcolor}
\usepackage{epstopdf}
\def\BibTeX{{\rm B\kern-.05em{\sc i\kern-.025em b}\kern-.08em
    T\kern-.1667em\lower.7ex\hbox{E}\kern-.125emX}}

\begin{document}
	\bstctlcite{IEEEexample:BSTcontrol}
	
\title{
	{\fontsize{22 pt}{\baselineskip}\selectfont Rate Splitting Multiple Access for Joint Communication and Sensing Systems with Unmanned Aerial Vehicles}}

\author{
  Yuwei~Li,
  Wanli~Ni,
  Hui~Tian,
  Meihui~Hua,
  and~Shaoshuai~Fan \\
   State Key Laboratory of Networking and Switching Technology \\
   Beijing University of Posts and Telecommunications, Beijing, China \\
   Email: \{yuweili, charleswall, tianhui, huameihui, fanss\}@bupt.edu.cn
}

\maketitle

\begin{abstract}
This paper investigates the problem of resource allocation for joint communication and radar sensing system on rate-splitting multiple access (RSMA) based unmanned aerial vehicle (UAV) system. UAV simultaneously communicates with multiple users and probes signals to targets of interest to exploit cooperative sensing ability and achieve substantial gains in size, cost and power consumption. By virtue of using linearly precoded rate splitting at the transmitter and successive interference cancellation at the receivers, RSMA is introduced as a promising paradigm to manage interference as well as enhance spectrum and energy efficiency. To maximize the energy efficiency of UAV networks, the deployment location and the beamforming matrix are jointly optimized under the constraints of power budget, transmission rate and approximation error. To solve the formulated non-convex problem efficiently, we decompose it into the UAV deployment subproblem and the beamforming optimization subproblem. Then, we invoke the successive convex approximation and difference-of-convex programming as well as Dinkelbach methods to transform the intractable subproblems into convex ones at each iteration. Next, an alternating algorithm is designed to solve the non-linear and non-convex problem in an efficient manner, while the corresponding complexity is analyzed as well. Finally, simulation results reveal that proposed algorithm with RSMA is superior to orthogonal multiple access and power-domain non-orthogonal multiple access in terms of power consumption and energy efficiency.
\end{abstract}
\section{Introduction}
Joint communication and sensing has recently attracted substantial attention \cite{luong2021radio}, which can be attributed to increased spectral utilization as well as reduced size, cost and power assumption due to sharing hardware and signal processing modules \cite{choi2016millimeter}. Meanwhile, unmanned aerial vehicles (UAVs) have been extensively employed as cost-effective aerial platforms to provide users with enhanced mobile service in data-demanding and emergency scenarios \cite{zeng2019accessing}.
\let\thefootnote\relax\footnotetext{This work was supported by the National Nature Science Foundation of China under Grant 61790553.}
On the one hand, sensing ability undertakes an important role in guaranteeing safe UAVs operation and air traffic management. Besides, UAVs can be leveraged as aerial sensing platforms to collect information \cite{ni2020optimal}, which significantly benefits from radar's ability to achieve contactless and privacy-preserving detection \cite{wu2021comprehensive}. On the other hand, the inevitable interference among radar and communication users requires effective interference management schemes. Moreover, UAV is restricted to its limited battery and endurance time, so energy-efficient design is much more involved.

By partly decoding the interference and partly treating it as noise, rate-splitting multiple access (RSMA) is viewed as a powerful framework to enhance spectrum and energy efficiency \cite{clerckx2016rate}. In downlink RSMA, the message intended for each user is divided into common and private parts which are simultaneously transmitted with superposition coding. The common message is decoded by all or a subset of users and removed before decoding the private message. The private message is dedicated to a specific user. The split of common and private messages can be flexibly adjusted so as to partly treating the interference as noise as presented in \cite{mao2019rate}. However, the characteristics of UAV present new challenges for RSMA. Specifically, compared to its terrestrial counterpart, UAV features in its high mobility and flexibility, which adds additional degree of freedom for system design \cite{li2018uav}. Finite battery onboard also requires higher energy utilization on the premise of guaranteeing communication quality.

Recently, some significant efforts have been devoted to UAV, RSMA, and joint sensing and communication systems. For example, \cite{xu2020rate} investigated beamforming design for joint radar and communication system, and testified RSMA could strike a better balance between communication rate and radar approximation error. \cite{chen2020performance} proposed a joint sensing and communication network where drone swarms simultaneously conduct radar sensing and data fusion communication by adopting beam sharing scheme. In \cite{rahmati2019energy}, RSMA was applied in cellular-connected UAV networks and it could outperform the non-orthogonal multiple access (NOMA) scheme in terms of energy efficiency. The authors of \cite{yu2019efficient} adopted rate-splitting to assist UAV in relaying data from computation-intensive devices, where the quality of services could be significantly improved. More detailedly, a comprehensive survey of multiple access schemes employed for UAV networks was presented in \cite{jaafar2020multiple}. However, the aforementioned works either were limited to RSMA for terrestrial base station (BS) or neglected the joint design of communication and sensing.

Motivated by the promising advantages and potential challenges of UAV, RSMA, and joint sensing and communication, it is legitimate to integrate them together to enhance cooperative sensing ability and network efficiency.
To the best of our knowledge, this is the first work implementing rate splitting for UAV-assisted joint communication and sensing systems.
The main contributions of this work can be summarized as follows:
1) we propose a novel framework of UAV-enabled sensing and communication system. To maximize the energy efficiency of the system, we jointly optimize the UAV location, the transmit beamforming and rate allocation of the transmitter under the constraints of transmit power budget, quality-of-service (QoS) requirement of users and radar approximation error;
2) we design an iterative algorithm to solve the formulated non-convex problem by addressing the UAV deployment subproblem and beamforming design subproblem iteratively;
3) we conduct numerical simulations to show that our proposed algorithm enhances the energy efficiency compared to orthogonal multiple access (OMA) and NOMA, which is beneficial for energy-constraint UAV to prolong its flight duration.

The rest of this paper is organized as follows. The system model and problem formulation are described in Section II. Then, the designed algorithm is provided in Section III. Next, numerical simulations are conducted in Section IV, which is followed by the conclusion in Section V.

\section{System Model and Problem Formulation}
\subsection{System Model}
Consider a downlink multi-antenna system shared by communication and radar detection in mmWave communication environment, which can simultaneously transmit probing signals to the targets located at the angles of interest and communicate with downlink users. The transmit antenna structure at the UAV is composed of a uniform linear array (ULA) with \emph{M} identical antenna elements placed horizontally. Each user is equipped with one antenna.

We assume that the UAV, with the projection points $z=(m,n)$ on the \emph{xy}-plane, is assigned a set $\mathcal{K}$ of ground users, indexed by $\{1,2...K\}$. As depicted in Fig. 1 , UAV undertakes the role of RSMA-assisted BS as well as a collocated Multiple Input Multiple Output (MIMO) radar in our scenario.

The message ${A_k}$ of user \emph{k} is split into common part ${A_{c,k}}$ and private part ${A_{p,k}}$. The common part of all \emph{K} users $\{A_{c,1}...A_{c,K}\}$ is jointly encoded into the common stream ${b_{c}}$, while the private one is respectively encoded into private signal $\{b_{1}…b_{K}\}$. Then the data stream vector is later precoded using the precoder $\bm{x}= [\bm{x}_c,\bm{x}_1...\bm{x}_K]$, $\bm{x_c}$ $\in$ $ \mathbb{C}$${^{M \times 1}}$ therein. So the transmit signal is:
\begin{align}
s(t) = \bm{x}_c{b_c}(t) + \sum\nolimits_{k \in \mathcal{K}} {\bm{x}_k{b_k}(t)}
\end{align}
where ${b_c}(t)$ is the common stream of time index $t$, and ${b_k}(t)$ is the private stream of user $k$. Based on \cite{stoica2007probing}, (1) meets the form of MIMO radar probing signal model and therefore is qualified for detecting task.\par
As presented in \cite{akdeniz2014millimeter}, the gains of NLoS paths are typically 20 dB weaker than that of LoS path in mmWave channels. Taking the mmWave’s characteristic into account \cite{rupasinghe2018non}, we assume a single LoS path for the mmWave channel. The channel vector is given as:
\begin{align}
{h_k} = \sqrt M \frac{{{\alpha _k}\bm{a}({\theta _k})}}{{PL{{(\sqrt {{H^2} + \parallel z - {z_k}{\parallel ^{^2}}} )}^{\frac{1}{2}}}}}
\end{align}
where $\bm{a}({\theta _k})$ is the path gain following standard complex Gaussian distribution, ${\theta _k}$ is the angle-of-departure (AoD) of the LoS path. We adopt the path loss model in $\rm{PL}(X)=1+X^{\gamma}$, where $\gamma$ is the path loss exponent, and $X$ is the Euclid distance of the LoS path between UAV and user $k$. \\
In (2), the array steering vector $\bm{a}({\theta _k})$ is:
\begin{align}
\bm{a}({\theta _k})=\frac{1}{{\sqrt M }}{[1,{e^{j2\pi \frac{D}{\lambda }\sin ({\theta _k})}}...{e^{j2\pi \frac{D}{\lambda }(M - 1)\sin ({\theta _k})}}]^T}
\end{align}
where $D$ is antenna element spacing along ULA and $\lambda$ is the wavelength of the carrier frequency.\par
According to (1)(2), the received signal at the kth user is:
\begin{align}
{y_k}(t) &= \bm{h}_k^H s(t)+ {n_k}(t) \nonumber
\\&= \bm{h}_k^H \bm{x}_c{b_c}(t) + \bm{h}_k^H\sum\nolimits_{k \in \mathcal{K}} {\bm{x}_k} {b_k}(t) + {n_k}(t)
\end{align}
where ${n_k (t)}$ is the complex Gaussian random variable with mean $0$ and variance ${\sigma _n}^2$.

\begin{figure} [!t]
	\centering
	\includegraphics[width=3.5 in]{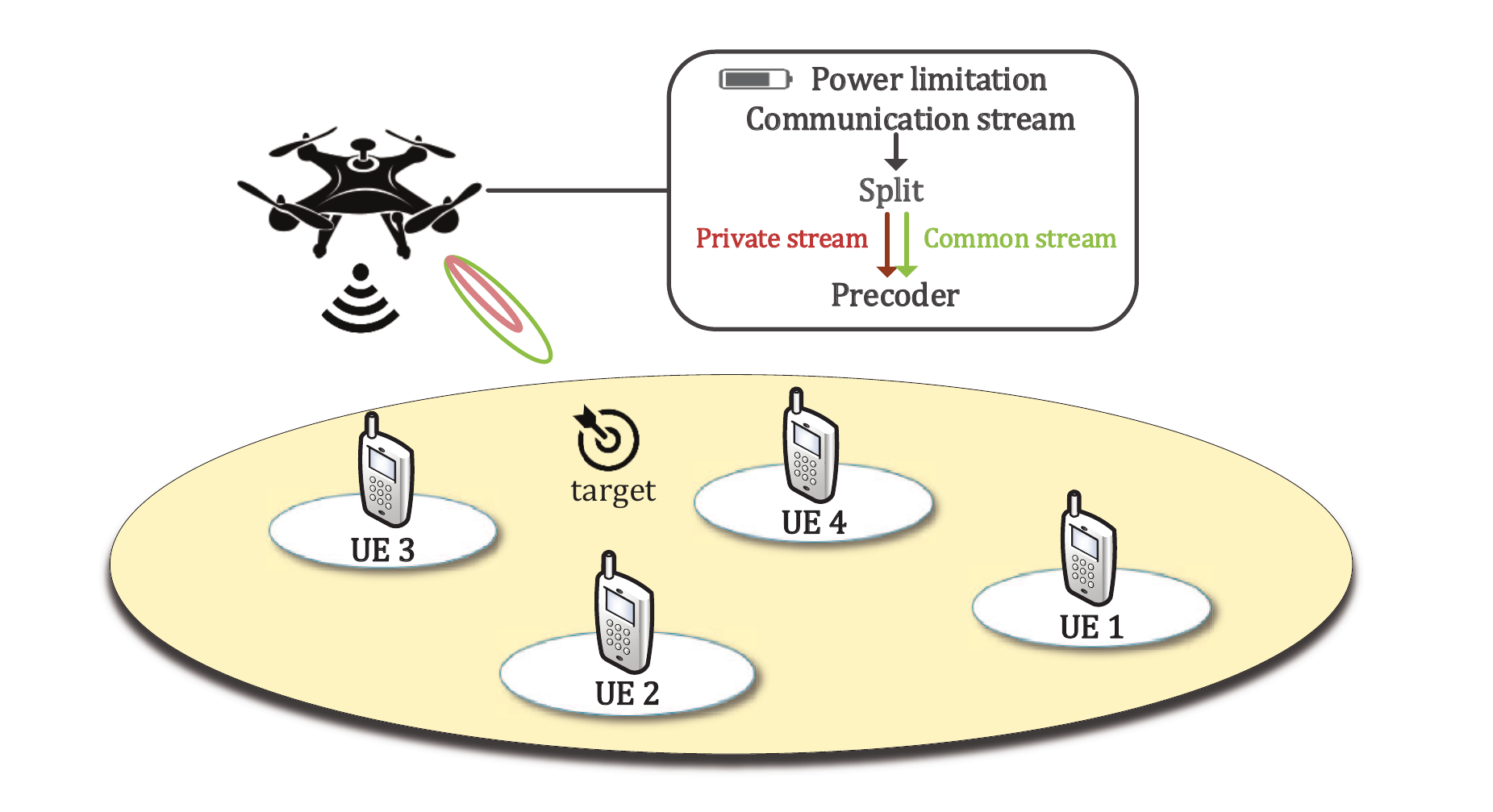}
	\caption{An illustration of UAV-enabled detection and communication system, where a radar is deployed on the UAV to complete the target detection task and RSMA technique is adopted to serve multiple ground users.}
	\label{system_model}
\end{figure}

At each user, the common stream is decoded firstly. Interference from the common stream is removed using successive interference cancellation \cite{mao2019rate}. Later, private part of the user's message is decoded, while treating private parts of other users as noise. So the SINR for decoding ${b_c}$ at user $k$ is:
\begin{align}
{r_k^c}(\bm{x}) = \frac{{|\bm{h}_k^H\bm{x}_c{|^2}}}{{\sum\nolimits_{j \in \mathcal{K}} {|\bm{h}_k^H\bm{x}_j{|^2} + \sigma _n^2} }},\forall k \in \mathcal{K}
\end{align}

Based on (5), the achievable rates of $s_c$ at user $k$ is $R_{_k}^c(\bm{x}) = {\log _2}(1 + r_{_k}^c(\bm{x}))$. After successfully decoding the common stream, and cancelling its contribution from the received signal, each user can decode its own private message with the interference of other private messages. The respective SINR for decoding $b_k$ at user $k$ is:
\begin{align}
{r_k}(\bm{x}) = \frac{{|\bm{h}_k^H\bm{x}_k{|^2}}}{{\sum\nolimits_{j \in K,j \ne k} {|\bm{h}_k^H\bm{x}_j{|^2} + \sigma _n^2} }},\forall k \in \mathcal{K}
\end{align}

The rates of $s_k$ at user $k$ is ${R_k}(\bm{x}) = {\log _2}(1 + r_k(\bm{x}))$. To guarantee that all users can decode the common stream, we have ${R_c}(\bm{x}) = \min \{ R_c^1(\bm{x})...R_c^K(\bm{x})\}$. As $R_c(\bm{x})$ is shared by $K$ users, we have:
\begin{align}
\sum\nolimits_{k \in \mathcal{K}} {{\beta _k} = {R_c}(\bm{x})}
\end{align}
where $\beta _k$ is the common transmission rate of user $k$ and $\bm{\beta}=\{\beta _1,\beta _2...\beta _K\}$ is the common rate allocation vector.\par
In this paper, we take $\mathrm{MIMO}$ radar design into account under the assumption of known target location. Based on \cite{stoica2007probing}, we aim at choosing $\bm{x}$ such that it approximates the desired beam pattern meeting the requirement of mean square error threshold. Consider there are $L$ targets of interest, to guarantee the performance for radar, there is:
\begin{align}
\sum\nolimits_{l = 1}^L {|\bm{a}^H({\theta _l})\bm{x}\bm{x}^H\bm{a}({\theta _l}) - {\zeta _l}{|^2}}  < \delta
\end{align}
where $\bm{a}({\theta _l}) = {[1,{e^{j2\pi \frac{D}{\lambda }sin({\theta _l})}}...{e^{j2\pi \frac{D}{\lambda }(M - 1)sin({\theta _l})}}]^T}$ is transmit steering vector, and ${\{\zeta _l\}}$ are constraint meant for the predetermined levels of radar detection at interesting grids.
In this study, we optimize the beamforming matrix $\bm{x}$, the location of UAV $\bm{z}$ and the common rate vector $\bm{\beta}$ to maximize the energy efficiency of UAV, which is given as:
\begin{align}
EE = \frac{{\sum\nolimits_{k = 1}^K {{\beta _k}}  + R_k^s(\bm{x})}}{{tr(\bm{x}\bm{x}^H)+ {P_{\rm{hov}}} + {P_{\rm{cir}}}}}
\end{align}
where $P_{\rm{hov}}$ is fixed energy consumption for hovering, and $tr(\bm{xx}^H)$ denotes antennas energy consumption antennas for joint communication and radar sensing. $P_{\rm{cir}}$ is circuit power consumption satisfying $P_{\rm{cir}}= MP_{\rm{dyn}}+P_{\rm{sta}}$, where $P_{\rm{dyn}}$ is dynamic power consumption of one active radio frequency chain and $P_{\rm{sta}}$ is static power consumption of cooling system.
\subsection{Problem Formulation}

As considered above, our goal is to optimize the beamforming matrix, the location of UAV and the common rate allocation vector so as to maximize the total energy efficiency of UAV. Power budget, Quality of service (QoS) requirements and beampattern design for radar are considered in constraints. The transmit beampattern problem for radar is restricted to approximation error for certain desired levels at a number of points in this paper. Mathematically, the formulated optimization can be written as:
\begin{subequations}
\begin{align}
\mathop {\max }\limits_{\bm{x,\beta ,z}}\quad& \frac{{\sum\nolimits_{k = 1}^K {{\beta _k} + R_k^s(\bm{x})} }}{{tr(\bm{xx}^H) + {\rm{P}_{hov}} + {\rm{P}_{cir}}}}\\
{\rm{s.t.}}\quad &\ \sum\nolimits_{{k^{'}} \in \mathcal{K}} {{\beta _{{k^{'}}}} \le R_k^c} (\bm{x}),\forall k \in \mathcal{K}\\
{}&{\beta_k}  \succeq 0 , \forall k \in \mathcal{K}\\
&tr(\bm{xx}^H) \le {P_{max}}\\
&\sum\limits_{l = 1}^L {|\bm{a}^H({\theta _l})\bm{x}\bm{x}^H\bm{a}({\theta _l}) - {\zeta _l}{|^2}}  < \delta  \\
&{\beta _k} + R_k^s(\bm{x}) \ge R_k ,\forall k \in \mathcal{K}
\end{align}
\end{subequations}
where $\rm{P}_{max}$ is the total power budget for communication and detection. Constraint (10b) indicates that the received common rates of all users must be less than the achievable common rate of any user. Constraint (10c) ensures that the assigned common rate of any user is non-negative. Constraint (10d) presents the maximum power constraint for detection and communication. Constraint (10e) guarantees us, in terms of detection, the mean square error between desired and realized beampattern are less than threshold. Constraint (10f) meets the QoS rate requirement of each user.
\section{Iterative Algorithm}
In this section, we propose an iterative algorithm with low complexity to solve the energy efficiency optimization problem (10). By decomposing (10) into the UAV deployment subproblem and beamforming subproblem, We adopt the successive convex approximation (SCA) and Dinkelbach method to alternatively optimize UAV location, transmit beamforming matrix and rate allocation vector.
\subsection{Location Optimization}
Given the fixed transmit beamforming matrix $\bm{x}$ and common rate allocation vector $\bm{\beta}$, the total power consumption of UAV is fixed. Then, the optimization problem is reformulated into private rate maximization problem as (11):
\begin{subequations}
\begin{align}
\mathop {\max }\limits_{\bm{z,f}}\quad & \sum\limits_{k = 1}^K {{{\log }_2}(1 + {f_k})}\\
{\rm{s.t.}}\quad &\frac{{|\bm{h}_k^H\bm{x}_c{|^2}}}{{\sum\nolimits_{j \in \mathcal{K}} {|\bm{h}_k^H\bm{x}_j{|^2} + \sigma _n^2} }} \ge {2^{{W_c}}} - 1 , \forall k \in \mathcal{K}\\
&\frac{{|\bm{h}_k^H\bm{x}_k{|^2}}}{{\sum\nolimits_{j \in \mathcal{K},j \ne k} {|\bm{h}_k^H\bm{x}_j{|^2} + \sigma _n^2} }}\ge {f_k} , \forall k \in \mathcal{K}\\
&{f_k} \ge {2^{R_k^{th} - {\beta _k}}} - 1 , \forall k \in \mathcal{K}\\
&\sum\limits_{l = 1}^L {|\bm{a}^H({\theta _l})\bm{x}\bm{x}^H\bm{a}({\theta _l}) - {\zeta _l}{|^2}}  < \delta  \\
&f_k \geq 0 , \forall k \in \mathcal{K}
\end{align}
\end{subequations}
where $f_k$ is a slack vector representing the transmission rate of private stream at the receiver side and respects constraint (11c). In (11b), ${W_c} = \sum\nolimits_{k = 1}^K {{\beta _k}}$ is the fixed common rate of all users. The equivalence between (11) and (10) is ensured only when constraint (11c) holds with equality at optimum.\par
Although $\bm{x}$ is fixed, according to (2), $\bm{h}_k$ is not convex with respect to the hight of UAV, namely $\bm{z}$ in (11).To handle this, we further introduce a auxiliary variable $\gamma_k$, ${\gamma _k} > 0, \forall k \in \mathcal{K}$ representing the interference plus noise at user $k$. Constraint (11c)  can be rewritten as:
\begin{subequations}
\begin{align}
&|\bm{h}_k^H\bm{x}_k{|^2} \ge {f_k}{\gamma _k},\forall k \in \mathcal{K}\\
&{\gamma _k} \ge \sum\nolimits_{j \in ,j \ne k} {|\bm{h}_k^H\bm{x}_j{|^2} + \sigma _n^2},\forall k \in \mathcal{K}
\end{align}
\end{subequations}

Then we opt for successive convex approximation (SCA) technique, where in each iteration, the left-hand side of (12a) is replaced by its concave lower bound at a given UAV location denoted by $\bm{z^r}$, with $r$ designating the $r$th iteration. As all convex functions are globally lower-bounded by its first-order Taylor expansion at any point, following the approach in \cite{jaafar2020downlink}, the left-hand side of (12a) can be approximated by:
\begin{align}
\omega {(\bm{z})_j} = \bm{A}_j^r(\bm{z} - {\bm{z}^r}) + W_j^r
\end{align}
\begin{align}
\bm{A}_j^r = \frac{{{{\log }_2}({\text{e}})\parallel {\bf{1}}{{\bf{x}}_j}{\parallel ^2}}}{{{\text{(}}\sum\nolimits_{i \in K,i \ne j} {\parallel {\bf{1}}{{\bf{x}}_i}{\parallel ^2}} {\text{ +  }}{{({\text{d}}_j^r{\text{)}}}^2}{\text{)(}}\sum\nolimits_{i \in K} {\parallel {\bf{1}}{{\bf{x}}_i}{\parallel ^2}} {\text{ +  }}{{({\text{d}}_j^r{\text{)}}}^2})}}
\end{align}
\begin{align}
W_j^r = \frac{{{{\log }_2}({\text{e}})\parallel {\bf{1}}{{\bf{x}}_j}{\parallel ^2}}}{{{\text{(}}\sum\nolimits_{i \in K,i \ne j} {\parallel {\bf{1}}{{\bf{x}}_i}{\parallel ^2}} {\text{ +  }}{{({\text{d}}_j^r{\text{)}}}^2}{\text{)(}}\sum\nolimits_{i \in K} {\parallel {\bf{1}}{{\bf{x}}_i}{\parallel ^2}} {\text{ +  }}{{({\text{d}}_j^r{\text{)}}}^2})}}
\end{align}
where $\bm{A}_j^r$ is and $W_j^r$ are coefficients of Taylor expansion, and ${\text{d}}_j^r = \parallel {{\mathbf{z}}^r} - {{\mathbf{z}}}{\parallel ^2}$.
The right side of (11c) can be rewritten as a DC function \cite{yang2020energy}. According to the above derivations, constraint (11c) of location optimization problem (11) can be replaced by:
\begin{align}
\omega {(\bm{z})_k} \ge &\frac{1}{2}({f_k} + {\gamma _k})^2- \frac{1}{2}(f{_k^r{}^2} + \gamma {_k^r{}^2})\nonumber \\
& - f_k^r({f_k} - f_k^r)- \gamma _k^r({\gamma _k} - \gamma _k^r)\nonumber \\
{\gamma _k} \ge &\sum\nolimits_{j \in \mathcal{K},j \ne k} \omega {(\bm{z})}_j + \sigma _n^2,\forall k \in \mathcal{K}
\end{align}

As for constraint (11b), we introduce variable $\varepsilon_k$ representing interference plus noise at each user $k$ to decode its public stream. Constraint (11b) is equivalent to:
\begin{subequations}
\begin{align}
\label{17a}
&|\bm{h}_k^H\bm{x}_c{|^2} \ge ({2^{{W_c}}} - 1){\varepsilon _k},\forall k \in \mathcal{K}\\
\label{17b}
&{\varepsilon _k} \ge \sum\nolimits_{j \in \mathcal{K}} \omega (\bm{z})_j + \sigma _n^2,\forall k \in \mathcal{K}
\end{align}
\end{subequations}
where the left side of (\ref{17a}) can be lower-bounded by $\omega{(\bm{z})_c} = \bm{A}_c^r(\bm{z} -{\bm{z}^r}) + W_c^r $. And $\bm{A}_c^r$ and $W_c^r$ are coefficients of Taylor expansion. Then, similar to the transition of constraint (11c), we plug the approximation into (17a) and constraint (11b) can be translated into:
\begin{align}
&\omega {(\bm{z})_c} \ge ({2^{{W_c}}} - 1){{\varepsilon _k}},\forall k \in \mathcal{K}\nonumber \\
&{\varepsilon_k} \ge \sum\nolimits_{j \in \mathcal{K}} \omega {{(\bm{z})}_j} + \sigma _n^2,\forall k \in \mathcal{K}
\end{align}

As for (11e), to make it more tractable, we derive a convex upper bound for the left side. Denote the left side of (11e) as $\bm{\mu _l}$ and its estimation at $\bm{z^r}$ can be expressed as :
\begin{align}
\sum\limits_{l = 1}^L \bm{W}_l^r + \bm{A}_l^r(\bm{z} - {\bm{z}^r})\le \delta
\end{align}
where $\bm{A}_l^r$ denotes the gradient of $\bm{\mu}_l$ with respect to $\bm{z}$ at $\bm{z}^r$ and $\bm{W}_l^r $ denotes $\bm{\mu} _l$ at $\bm{z}^r$.

Motivated by aforementioned approximations, problem (11) is equivalently transformed into:
\begin{subequations}
\begin{align}
\mathop {\max }\limits_{\bm{z,f,\gamma ,\varepsilon }}\quad & \sum\limits_{k = 1}^K {{{\log }_2}(1 + {f_k})}\\
{\rm{s.t.}}\quad &(16), (18), (11d), (19)\\
&{f_k} \ge 0,{\gamma _k} \ge 0,{\varepsilon _k} \ge 0,{\theta _k} \ge 0,\forall k \in \mathcal{K}
\end{align}
\end{subequations}
The above problem is convex and can be solved using CVX \cite{grant2009cvx}. The detailed process of using the SCA method to solve problem (11) is given in Algorithm 1.
\begin{algorithm}[t]
\caption{Iterative Optimization for Problem (11)}
\label{algorithm 1}
\begin{algorithmic}[1]
\renewcommand{\algorithmicrequire}{\textbf{Initialize}}
\STATE \textbf{Initialize} $(\bm{\beta}^0, \bm{x}^0)$. Set iteration number $t$ and the tolerance error $\epsilon_1 $.
\REPEAT
\STATE Solve convex problem (20) under given beamforming matrix and rate allocation vector for UAV location.
\STATE Denote the optimal solution of (20) by $\bm{z}^t$.
\STATE Set $t = t+1$;
\UNTIL the objective value (11) converges.
\end{algorithmic}
\end{algorithm}
\subsection{Beamforming and Rate Allocation Optimization}
Given fixed UAV location $\bm{z}$, problem (10) can be reformulated with the SCA-based algorithm. Problem (10) becomes:
\begin{subequations}
\begin{align}
\mathop {\max }\limits_{\bm{\beta ,x,f\gamma ,\varepsilon ,\theta }}\quad & \frac{{\sum\nolimits_{k = 1}^K {{\beta _k} + {{\log }_2}(1 + {f_k})} }}{{tr(\bm{x}\bm{x}^H) + {P_{\rm{hov}}} + {P_{\rm{cir}}}}}\\
{\rm{s.t. }}\quad&{}{\gamma _k} \ge \sum\nolimits_{j \in \mathcal{K},j \ne k} {|\bm{h}_k^H\bm{x}_j{|^2} + \sigma _n^2} \\
&|\bm{h}_k^H\bm{x}_c{|^2} \ge {\theta _k}{\varepsilon _k},\forall k \in \mathcal{K}\\
&|\bm{h}_k^H\bm{x}_k{|^2} \ge {f_k}{\gamma _k},\forall k \in \mathcal{K}\\
&{\varepsilon _k} \ge \sum\nolimits_{j \in \mathcal{K}} {|\bm{h}_k^H\bm{x}_j{|^2} + \sigma _n^2},\forall k \in \mathcal{K} \\
&{f_k} \ge {2^{R_k^{th} - {\beta _k}}} - 1,\forall k \in \mathcal{K}\\
&\ \sum\nolimits_{{k^{'}} \in \mathcal{K}} {{\beta _{{k^{'}}}} \le {{{\log }_2}(1 + {\theta_k})}},\forall k \in \mathcal{K}\\
&tr(\bm{xx}^H) \le {P_{max}}\\
&\sum\limits_{l = 1}^L {|\bm{a}^H({\theta _l})\bm{x}\bm{x}^H\bm{a}({\theta _l}) - {\zeta _l}{|^2}}  < \delta
\end{align}
\end{subequations}
where $\theta_k$ is a slack variable representing the transmission rate of public information.

According to \cite{yang2020energy}, the term $\bm{h}_k^H\bm{x}_k$ in constraint (21d) can be expressed as a real number through an arbitrary rotation to beamforming matrix $\bm{x}_c$. Therefore, (21d) is translated into:
\begin{align}
\mathcal{R}(\bm{h}_k^H\bm{x}_k) \ge &\sqrt {{f_k}^r{\gamma _k}^r}  + \frac{1}{2}\sqrt {\frac{{{f_k}^r}}{{{\gamma _k}^r}}} ({\gamma _k} - {\gamma _k}^r)\nonumber \\
& + \frac{1}{2}\sqrt {\frac{{{\gamma _k}^r}}{{{f_k}^r}}} ({f_k} - {f_k}^r),\forall k \in \mathcal{K}
\end{align}
The term $\bm{h}_k^H\bm{x}_c$ in constraint (21c) is composed of $K$ inequality constraints and thus cannot be translated like $\bm{h}_k^H\bm{x}_k$. The DC approximation is adopted to deal with the right side of (21c) and $|\bm{h}_k^H\bm{x}_c{|^2}$ is approximated by its first order Taylor expansion. Constraint (21c) is translated into:
\begin{align}
& \qquad \qquad \qquad 2\mathcal{R}(\bm{x}{_c^r{}}^H\bm{h}_k\bm{h}_k^H\bm{x}_c)-|\bm{h}_k^H\bm{x}_c^r{|^2} \ge\nonumber\\
 &\frac{1}{4}({({\varepsilon _k} + {\theta _k})^2} - ({\theta _k}^r - {\varepsilon _k}^r)({\theta _k} - {\varepsilon _k}) + {({\theta _k}^r - {\varepsilon _k}^r)^2}),\forall k
\end{align}

Like (19) derived above, we derive the convex upper bound for left side of (21i) at $\bm{x}^r$. Constraint (21i) is replaced by:
\begin{align}
\sum\limits_{l = 1}^L \bm{W}_l^r + \bm{A}_l^r(\bm{x} - {\bm{x}^r})\le \delta
\end{align}
According to \cite{15}, since the objective function is a concave function divided by a convex function and the feasible set is convex, the optimal solution of problem (21) is quasi-concave and can be tackled by the Dinkelbach method in \cite{15}.
We further introduce parameter $\tau $, the non-convex objective function (26a) is replaced by the following convex one:
\begin{align}
\varphi (\bm{\beta,x,f,\gamma ,\varepsilon,\theta }&|\tau )  =  tr(\bm{xx}^H) + {P_{\rm{hov}}} + {P_{\rm{cir}}} \nonumber\\
&- \tau (\sum\limits_{k = 1}^K {{\beta _k}}  + {\log _2}(1 + {f_k}))
\end{align}
Then, the non-convex problem (21) can be reformulated as
\begin{subequations}
\begin{align}
\mathop {\min }\limits_{\beta ,x,f,\gamma ,\varepsilon ,\theta }\quad & \varphi (\bm{\beta ,x,f,\gamma ,\varepsilon ,\theta }|\tau )\\
{\rm{s.t.}}\quad &{f_k} \ge 0,{\gamma _k} \ge 0,{\varepsilon _k} \ge 0,{\theta _k} \ge 0\\
 &(21b),(21e),(21f),(21g),(21h),\nonumber \\
&(21i),(22),(23),(24)
\end{align}
\end{subequations}
(26) is a convex problem and can be effectively solved by CVX in Matlab \cite{grant2009cvx}.
\begin{algorithm}[t]
\caption{Iterative Optimization for Problem (10)}
\label{algorithm 2}
\begin{algorithmic}[1]
\renewcommand{\algorithmicrequire}{\textbf{Initialize}}
\renewcommand{\algorithmicensure}{\textbf{Until}}
\STATE \textbf{Initialize} $(\bm{\beta}^0,\bm{x}^0,\bm{z}^0)$. Set iteration number as $t=1$ and the tolerance error $\epsilon_2$.
\REPEAT
\STATE Given ($\bm{\beta}^0,\bm{x}^0$), obtain ${\bm{z}^t}$ by solving (20) with Algorithm 1;
\STATE Given ${\bm{z}^t}$, obtain $(\bm{\beta}^t,\bm{x}^t)$ by solving (26) with CVX;
\STATE Set $t = t + 1$;
\UNTIL the objective value of (10) converges.
\STATE \textbf{Output} the converged solutions $\bm{\beta}^*$, $\bm{x}^*$ and $\bm{z}^*$.
\end{algorithmic}
\end{algorithm}
\subsection{Complexity Analysis}
In summary, our proposed iterative algorithm for solving (10) is given in Algorithm 2. At each iteration, the complexity of solving problem (10) is mainly determined by the complexity of solving (20) and (26).  The location optimization problem (11) consist of $(4K+1)$ constraints. Iterations number required by the SCA method is $\mathcal {O}(\sqrt{9K+1}\log_2(1/\epsilon_1))$ \cite{grant2009cvx}, where $\epsilon_1$ is the accuracy of the SCA method for solving problem (20). The complexity of solving problem (20) is $\mathcal {O} ({T_1^2}T_2)$, where $T_1=3K+2$ is the overall number of variables and $T_2=9K+1$ is the number of constraints. Therefore, complexity of problem (11) is $\mathcal {O}({K^{3.5}}\log_2(1/\epsilon_1))$. We adopt similar analysis and get the complexity for solving the beamforming problem (21) is $\mathcal {O}(TK^{3.5}\log_2(1/\epsilon_2))$, where $T$ is the number of iterations for solving problem (26) with Dinkelbach method and $\epsilon_2$ is the accuracy of the SCA method for solving problem (21). As a result, the total complexity of Algorithm 2 for solving problem (10) is $\mathcal {O}(SK^{3.5}\log_2(1/\epsilon_1)+STK^{3.5}\log_2(1/\epsilon_2))$, where $S$ is the number of iterations for Algorithm 2.
\section{Simulation Results}
In this section, we present the simulation results to illustrate the performance of the proposed algorithm. We suppose $K=4$ users uniformly distributed in a 50 m $\times$ 50 m square area with one radar target. We assume UAV employs a ULA with half-wavelength spacing, i.e., $\frac{D}{\lambda } = 0.5$ and $M = 8$ transmit antennas. For the UAV system, we compare the performance of the proposed RSMA system with NOMA and OMA. The simulation parameters are listed in Table I.\par

\begin{table}[h]
	\centering
	\caption{System Parameters}
	\begin{tabular}{cc}
		\hline
		\hline
		Parameters &Values\\ \hline
		Path loss exponent $\gamma$ & 2 \\
		Path gain, $\forall k $ & $a(\theta _k)\sim\mathcal {C}\mathcal {N}(0, 1)$ \\
		UAV height $H$ & 50m\\
		Noise power ${\sigma^2_n}$ & -50 dBm\\
		Fixed power consumption $\rm{P}_{hov}+\rm{P}_{cir}$ & 30dBm\\
		Transmit power budget $\rm{P}_{max}$ & 26 dBm\\
		approximation error ${\delta}$ & -20dB\\ \hline
		\hline
	\end{tabular}
\end{table}

\begin{figure} [t]
	\centering
	\includegraphics[width=2.6 in]{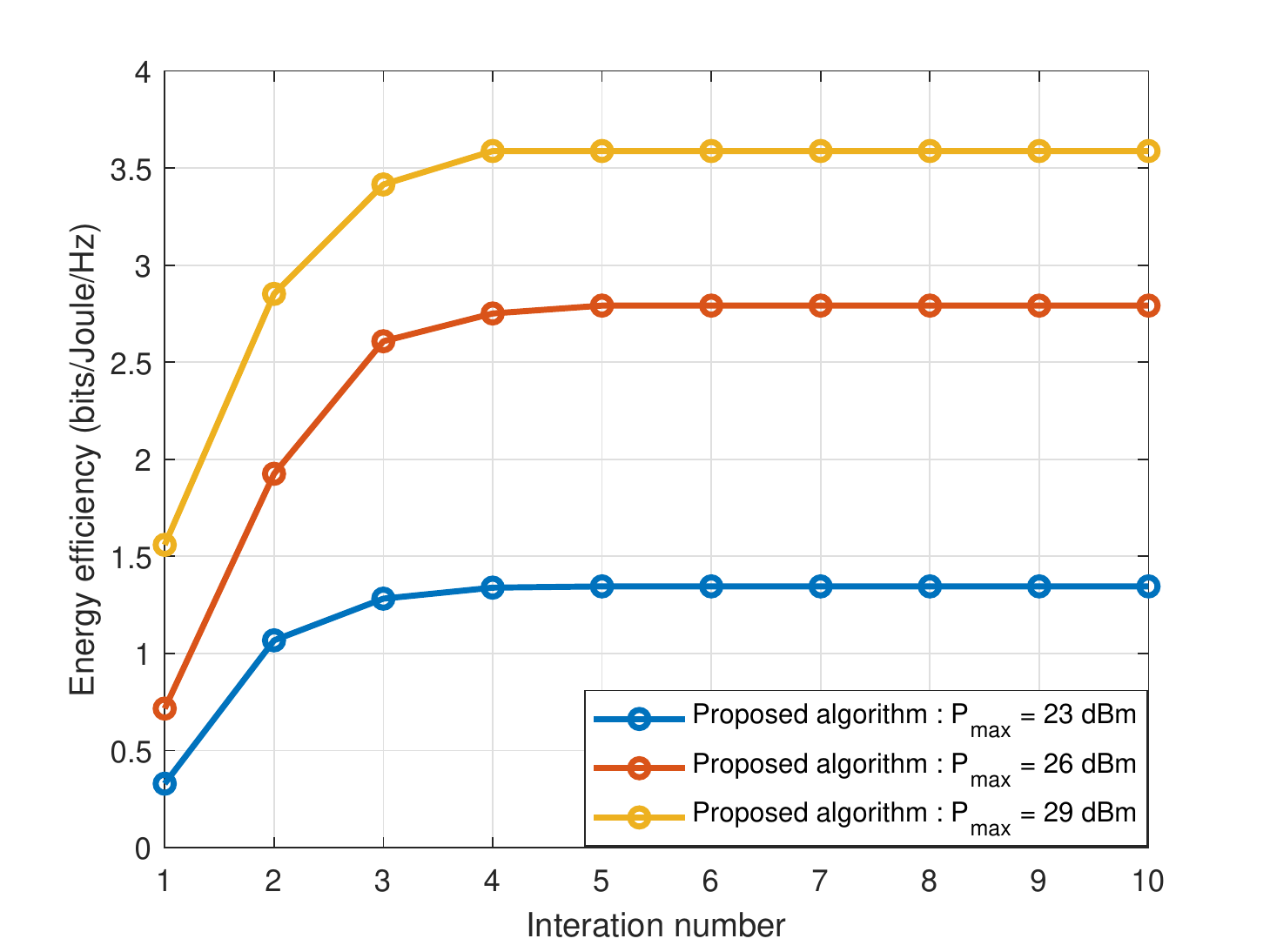}
	\caption{Convergence of the proposed algorithm with different power budget.}
\end{figure}

Fig. 2 shows the convergence of the proposed algorithm under different power budget. The final stable value increases monotonously with the power budget. Moreover, we can notice that the rising speed of energy efficiency goes down as the power budget increases. Besides, the final value at 26 dBm is twice as much as that of 23 dBm, but only 25$\%$ smaller than the value at 29 dBm. That's because the energy efficiency rises rapidly for all schemes when power budget increases from a relative low baseline. As the power budget rises, the increasing rate of energy efficiency slows down.

Fig. 3 shows the energy efficiency versus power budget. From this figure, RSMA reaches up to 12$\%$ and 24$\%$ gains in terms of energy efficiency compared to NOMA and OMA, respectively. Both RSMA and NOMA outperform OMA, owing to interference cancellation utilized by above two schemes. RSMA achieves the best performance counting on the fact that RSMA dynamically treats interference between radar and users as noise or as interference, while the NOMA only treats it as interference and users decode them all.

\begin{figure} [!t]
	\centering
	\includegraphics[width=2.6 in]{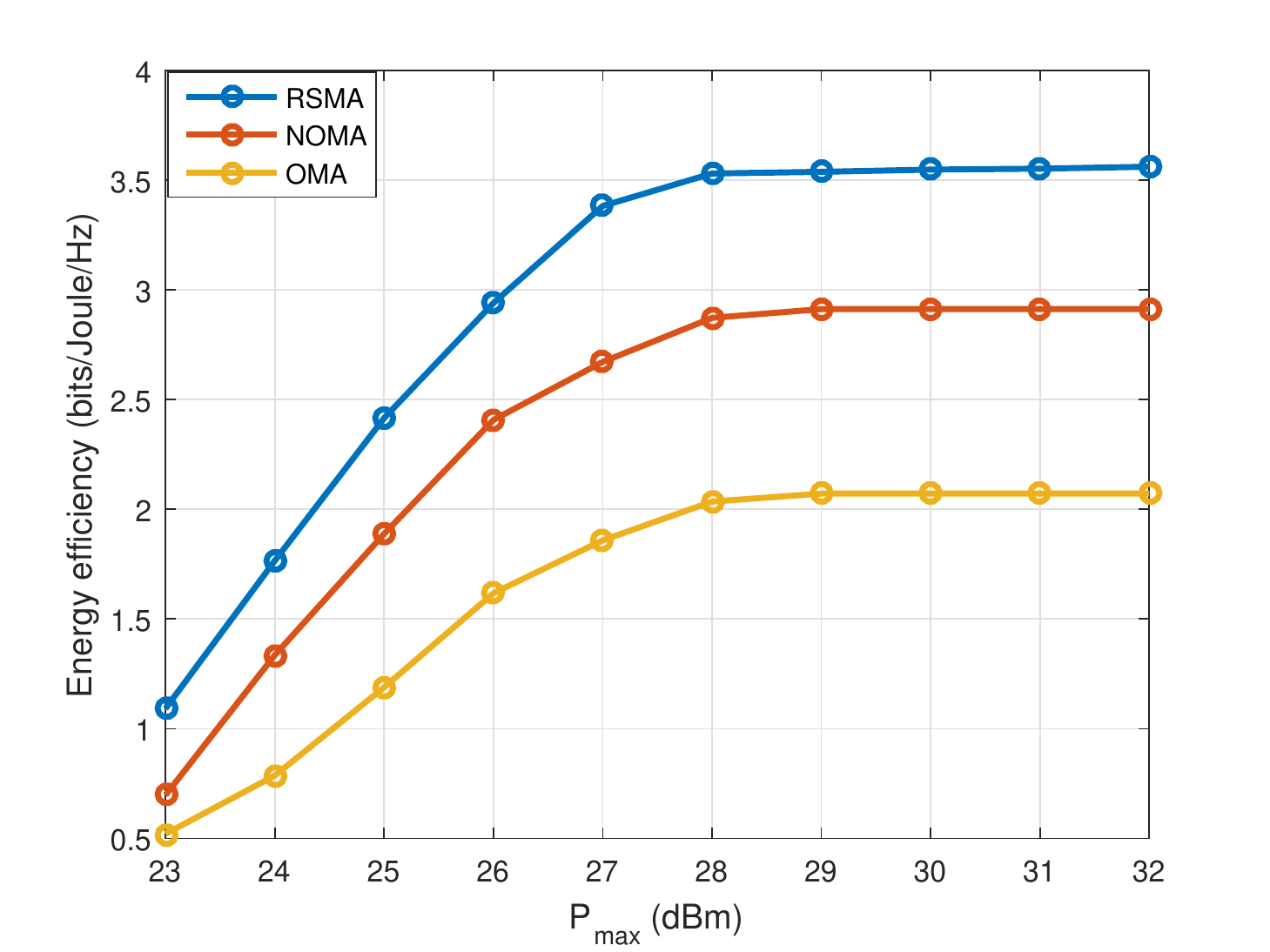}
	\caption{Energy efficiency versus power budget.}
\end{figure}

\begin{figure} [!t]
	\centering
	\includegraphics[width=2.6 in]{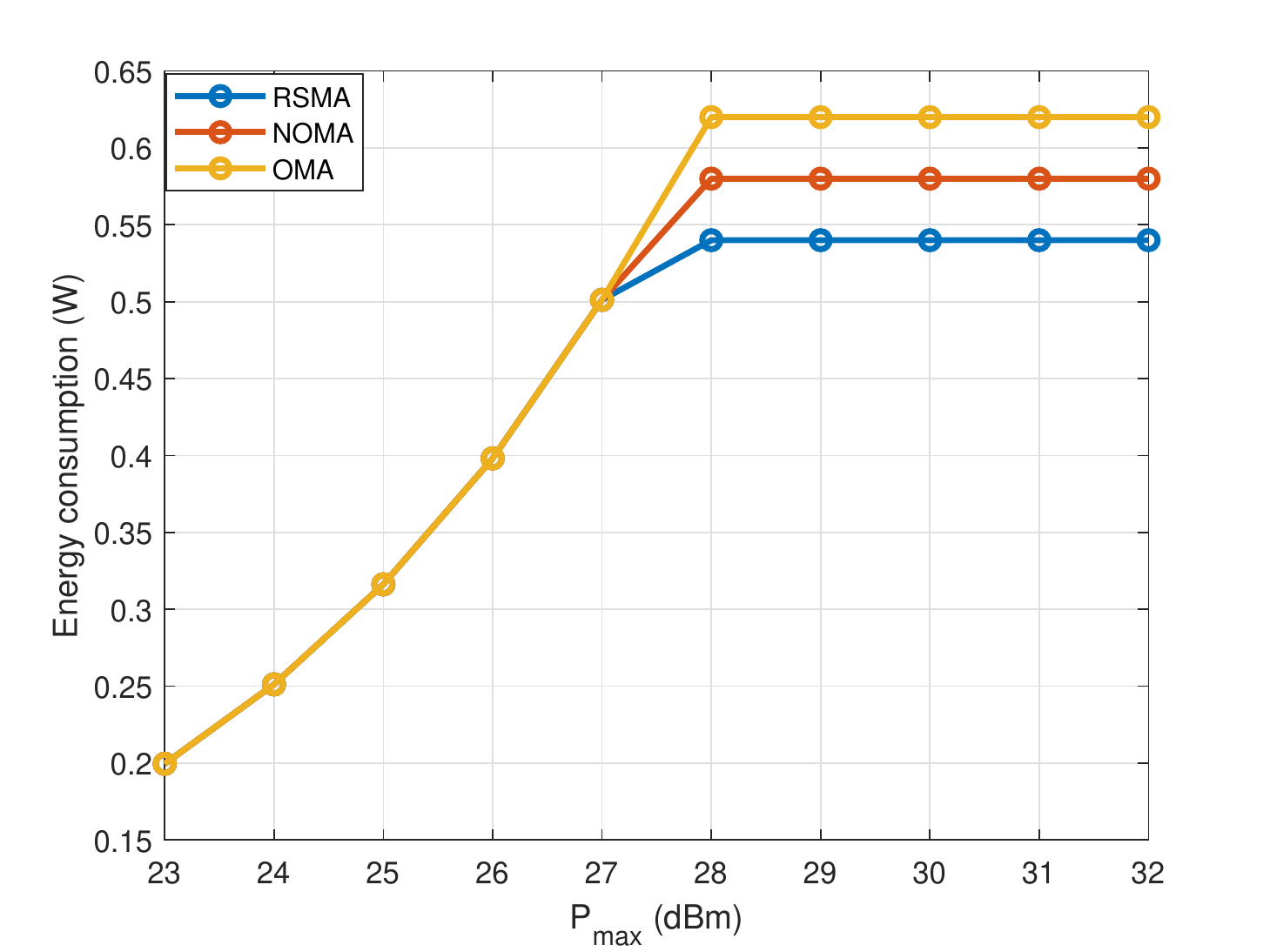}
	\caption{Energy consumption versus power budget.}
\end{figure}

Fig. 4 shows the energy consumption for detection and communication versus power budget. The energy consumption rises at first as the power budget increases from a relative low value. That illustrates power budget is the limiting factor to energy efficiency. And all three strategies make the most of energy by keeping the maximum energy consumption. When the power budget reaches 28 dBm, the value remains stable. This corresponds to the stable point in Fig. 3, which implicates that the energy eifficicy arrives at optimum at this point and further increase of power budget brings no more gain in system performance.Besides, the optimum energy consumption of RSMA is the lowest of all three strategies and NOMA consume less energy than OMA.


\section{Conclusion}
In this paper, we have considered resource allocation for RSMA for joint communication and sensing on UAV platform. Delpoyment of UAV, transmit beamforming and rate allocation are jointly optimized to maximize the energy efficiency of UAV while satisfying power budget, QoS requirements and radar approximation error. We have proposed an iterative algorithm with low complexity which iteritively optimize UAV location sub-problem and beamforming sub-problem. Numerical results have shown that the proposed scheme outperforms NOMA and OMA in terms of energy efficiency.

\bibliographystyle{IEEEtran}
\bibliography{mybib}

\end{document}